\documentstyle[psfig]{mn}
\def\H0{{\it H}$_0$}

\def\q0{{\it q}$_0$}

\begin{document}
\title[Large scale jets in AGN: multiwavelength mapping]
{Large scale jets in AGN: multiwavelength mapping}
\author[Celotti, Ghisellini and Chiaberge]
{Annalisa Celotti$^1$, Gabriele Ghisellini$^2$ and Marco Chiaberge$^1$\\
$^1$ SISSA, Via Beirut 2--4, I--34014 Trieste, Italy \\
$^2$ Osservatorio Astronomico di Brera, Via Bianchi 46, I--23807
Merate (Lc), Italy \\
E--mail: {\tt celotti@sissa.it}, {\tt gabriele@merate.mi.astro.it}, 
{\tt chiab@sissa.it}
}


\maketitle

\begin{abstract}
X--ray emission from large scale extragalactic jets is
likely to be due to inverse Compton scattering of relativistic
particles off seed photons of both the cosmic microwave background
field and the blazar nucleus. 
The first process dominates the observed high energy 
emission of large scale jets if the plasma is moving at
highly relativistic speeds and if the jet is aligned
with the line of sight, i.e. in powerful flat radio spectrum quasars.
The second process is relevant when the plasma is moving at
mildly bulk relativistic speeds, and can dominate the high energy
emission in misaligned sources, i.e. in radio--galaxies.
We show that this scenario satisfactorily accounts for the
spectral energy distribution detected by $Chandra$ from the jet and
core of PKS~0637--752.
\end{abstract}
\begin{keywords}
galaxies: active - galaxies: jets - galaxies: quasars: individual:
PKS~0637--752 - radiation mechanisms: non thermal
\end{keywords}

\section{Introduction}

The $Chandra$ X--ray Observatory (CXO) is providing us with unprecedented
high spatial (and spectral) resolution data on extended structures.
The very first observation of a radio--loud quasar, PKS~0637--752, revealed the
presence of an X-ray jet extending for $\sim$ 10 arcsec (Chartas et
al. 2000; Schwartz et al. 2000), and since then other X--ray jets have
been detected with high spatial resolution (Cen A, Pictor A, see CXO www page
{\tt http://chandra.harvard.edu/photo/cycle1.html}).

While the detection of an X--ray jet is not unprecedented (e.g.  M87,
Biretta, Stern \& Harris, 1991; 3C273, Harris \& Stern 1987; R\"oser
et al. 2000), it is interesting to notice that both radio--galaxies and
blazar--like sources (i.e.  with a jet oriented close to the line of
sight) have been observed. 
Models involving inverse Compton scattering
of the Cosmic Microwave Background (CMB) and nuclear hidden quasar
radiation to produce large scale X--rays have been proposed
(e.g. Brunetti, Setti \& Comastri 1997 in the case of emission
from the lobes).  
However it has been found
that the local broad band spectral energy distributions from the knots
in both PKS~0637--752 and M87, are not straightforward to interpret,
since the level of the X--ray emission is higher than what simple
models predicts (e.g. Chartas et al. 2000; Schwartz et al. 2000,
R\"oser et al. 2000).

Here we propose a scenario involving both highly and
mildly relativistically moving plasma in the jet and consider the
dominant radiation fields on which relativistic particles Compton
scatter. This can satisfactorily account for
enhanced large scale X--ray emission in all jetted sources, both
blazars and radio--galaxies.  We also consider the spatial
distribution of emission at different frequencies (Section 3). The
predictions are compared to the results on PKS~0637--752 in Section 4
and our conclusions drawn in Section 5.

\section{Mirrors in the jet}

Comparatively little is observationally known on the structure and
dynamics of extragalactic jets. 
Theoretically one would expect the existence of a 
velocity structure along the jet, from a faster
inner core to slower outer regions of the flow -- dissipation can be
produced by shocks in the relativistic plasma flow, or deceleration of
the jet material can be caused by obstacles in the jet or by the
interaction of the relativistic flow with the (steady) walls. 
However only recently observational evidence has been
accumulating in support of such possibility and on the relevant role
of the outer layers in contributing to the observed emission.  This
evidence includes the morphology and polarization structure of large
scale low power jets (Komissarov 1990, Laing 1993, 1999; 
Giovannini et al. 1999) as well as
apparent inconsistencies on the value of the relativistical beaming
parameters as inferred from the flux observed at large angles with the
jet direction and the statistic properties of beamed and parent
populations according to the unification schemes (Chiaberge et
al. 2000). These facts have been satisfactorily accounted for by
schematically consider the jet as constituted by two main components:
a slow, but still mildly relativistic, outer ``layer'' and a high
relativistically moving ``spine''. Strong support for such a fast
component even at large (arcsec) distances from the active nucleus
comes from the detected superluminal motion of M87 on kpc scales
(Biretta, Sparks \& Macchetto 1999).

Within this assumption on the jet structure, let us consider the
electromagnetic fields (magnetic and radiative) which can contribute
to the emission of relativistic electrons at different distances along
the jet, and thus determine the corresponding dominant cooling process
for relativistic particles.  Indeed, if dissipation occurs in knots at
large distances, the accelerated particles can cool by the synchrotron
and synchrotron self--Compton (SSC) processes, but also by scattering
any radiation produced externally to the knot itself. The relative
importance of the fields clearly depends on the bulk Lorentz factor of
the emitting regions in the jet,
i.e. the spine and the layer bulk Lorentz factors.

\subsection{Radiation fields}

Theoretical advances in recent years (especially due to high energy
$\gamma$--ray observations) have convinced us that the bulk of the
emission seen in blazars is produced in a rather well defined region
of the jet, which on one hand cannot be too compact (hence, close to
the jet apex) in order not to absorb $\gamma$--ray photons through
$\gamma$--$\gamma$ interactions, and on the other hand cannot be too
large in order to vary on short timescales (Ghisellini \& Madau 1996).
This region should be located at some hundreds of Schwarzschild radii
from the black hole, and produce radiation collimated in the beaming
cone of semi--aperture angle $\sim 1/\Gamma_{\rm in}$, where
$\Gamma_{\rm in}$ is its bulk Lorentz factor.  At larger distances,
the jet is therefore illuminated by this radiation: it turns out that
it can be observationally relevant as scattered radiation if the large
scale plasma has only mildly relativistic speeds, as supposed for the
jet layer.

In fact, let us assume that on the parsec and sub--parsec scale the
jet contains a blazar--like emitting region, moving with a bulk
Lorentz factor $\Gamma_{\rm in}$ at an angle $\theta$ with the line of
sight.  This region emits an intrinsic (comoving) synchrotron
luminosity $L^\prime_{\rm s,in}$.
Further out, at a distance $z$ from the jet apex, there is a knot
where dissipation occurs and electrons are accelerated and therefore
radiate.  If the bulk Lorentz factor $\Gamma_{\rm out}$ of this region
is significantly smaller than $\Gamma_{\rm in}$ this region will see
the nuclear (inner) flux enhanced by beaming, with a corresponding 
radiation energy density
\begin{equation}
U^\prime_{\rm in} \, \sim \, {L_{\rm s,in} \over 4 \pi z^2
c\delta_{\rm in}^4} {\Gamma^{2}_{\rm in} \over \Gamma^{2}_{\rm out}}
\end{equation}
where $\delta_{\rm in}\equiv \Gamma^{-1}_{\rm in}[1-\beta_{\rm
in}\cos(\theta)]^{-1}$ is the beaming or Doppler factor of the inner
jet, and $\beta_{\rm in}c$ is the corresponding velocity.

This external energy density must be compared with the local
magnetic ($U^\prime_{\rm B}$) and synchrotron radiation
($U^\prime_{\rm out}$) energy densities.
If the magnetic field is in equipartition with the locally produced
synchrotron radiation, we have $U^\prime_{\rm B}\sim U^\prime_{\rm out}$,
and the self--Compton emission is then of the same order. 
Alternatively, the field can be estimated by conservation of magnetic 
flux $L_{\rm B}$, which at large distances gives
\begin{equation}
U'_{\rm B} \, = \, {L_{\rm B} \over \pi \psi^2 z^2 c \Gamma^2_{\rm out} },
\end{equation}
where we assume that the emitting region located at $z$ has
a transverse dimension $R=\psi z$.
In general $\psi$ can be a function of $z$ if the jet is not
conical, especially at $>$kpc scale, where some recollimation
may occur (see e.g. Kaiser \& Alexander 1997). 
The synchrotron energy density can be written as
\begin{equation}
U^\prime_{\rm out} \, =\, 
{L_{\rm s,out} \over 4 \pi \psi^2 z^2 c \delta^4_{\rm out}}.
\end{equation}
where $\delta_{\rm out}$ is the beaming factor of the knot radiation.

\subsection{The quasi--isotropic fields}

Further fields which can provide seed photons for 
Compton scattering are those reprocessed in the Narrow Line Region
(NLR) and by any large scale dust structure surrounding the nucleus. 
We consider these two components as produced within $\sim$kpc 
scale and therefore they would be seen de--amplified by plasma 
moving along the jet on larger scales:
\begin{equation}
U^\prime_{\rm kpc} \sim U^\prime_{\rm NLR} + U^\prime_{\rm dust}
\sim 
{{f L_{\rm disk} +4\pi z_{\rm dust}^2 \sigma T_{\rm dust}^4} 
\over {4 \pi z^2 c}\,
\Gamma_{\rm out}^{2}};\, z>z_{\rm NLR}, z_{\rm dust}
\end{equation}
where $f L_{\rm disk}$ is the fraction of the disk emission
reprocessed by the NLR. Dust is assumed to re-emit as
a quasi-blackbody at temperature $T_{\rm dust}$ at a distance $z_{\rm dust}$.

Finally, the CMB radiation provides 
a uniformly distributed contribution to the radiation fields:
\begin{equation}
U^\prime_{\rm CMB} \, \sim \, a T_{\rm CMB}^4 (1+z)^4\,\Gamma_{\rm
out}^{2}.
\end{equation}
The inverse Compton process on the CMB photons is particularly relevant if 
the knot is moving relativistically, since in this case the CMB 
energy density as seen in the comoving frame is amplified by 
$\Gamma_{\rm out}^2$.



A possible contribution to the seed photons comes from the interplay 
between the spine and the layer of the jet, if both are active and 
relatively close.
Radiation produced in one zone can in fact be seen amplified 
(by beaming) by the other zone.
We have however verified that these contributions are less important than 
the other ones, at least for the range of parameters considered in this work.
We have also neglected any stellar contribution. 


\subsection{The relevant fields}

The relative importance of the radiation produced by
the blazar component and the local SSC emission is given by
\begin{equation}
{U^\prime_{\rm in} \over U^\prime_{\rm out}} \, = \,
{(\psi \Gamma_{\rm in})^2 \over \Gamma_{\rm out}^2}
 {L_{\rm s,in} \over L_{\rm s,out}}\, 
\left( {\delta_{\rm out} \over \delta_{\rm in} }\right)^4  \, =\,
{(\psi \Gamma_{\rm in})^2 \over \Gamma_{\rm out}^2} \, 
{L^\prime_{\rm s,in}\over L^\prime_{\rm s,out}}.
\end{equation}
%
%
%
For illustration, consider the two cases of 
$\Gamma_{\rm out}=\Gamma_{\rm in}$ (corresponding to a fast moving spine)
and $\Gamma_{\rm out}\sim 1$ (corresponding to a layer at rest
relative to the center):

i) $\Gamma_{\rm out}=\Gamma_{\rm in}$:
in this case 
$U^\prime_{\rm in}$ dominates if $L_{\rm s,in}>L_{\rm
s,out}/\psi^2$, i.e. if the observed power coming from the nucleus is
two orders of magnitude (for $\psi\sim 0.1$) greater than the power of
the knot; 

ii) $\Gamma_{\rm out}\sim 1$:
in this case, for $\psi \Gamma_{\rm in}$ of the order of unity,
$U^\prime_{\rm in}$ dominates if the power dissipated in the inner
jet is greater than the power dissipated at large scales,
i.e. if $L^\prime_{\rm s,in}>L^\prime_{\rm s,out}$.
  
We conclude that it is likely that some part of the outer
jet can produce -- depending on the relative velocities of the inner
and outer regions -- an inverse Compton spectrum dominated by seed
photons coming from the inner regions, for $L_{\rm s,in}> 10^2$--$10^4$
$L_{\rm s,out}$.


In order to fully compare all the contributions discussed above
to the Compton emission of the knot, in Fig. 1 we report  
the relevant energy densities as a function of the distance 
along the jet, for two values of the bulk Lorentz factors 
(schematically corresponding to a ``spine'' and ``layer'').
For semplicity, in Fig. 1 we also assumed that $\psi$
is constant (and equal to 0.1), i.e. a conical jet.
Clearly, if the jet progressively recollimates at large $z$,
the scaling of $U_{\rm B}$ and $U_{\rm s, out}$ changes
and their values increases relatively to the other energy densities.
\begin{figure}
\vskip -0.5 true cm
\centerline{\psfig{figure=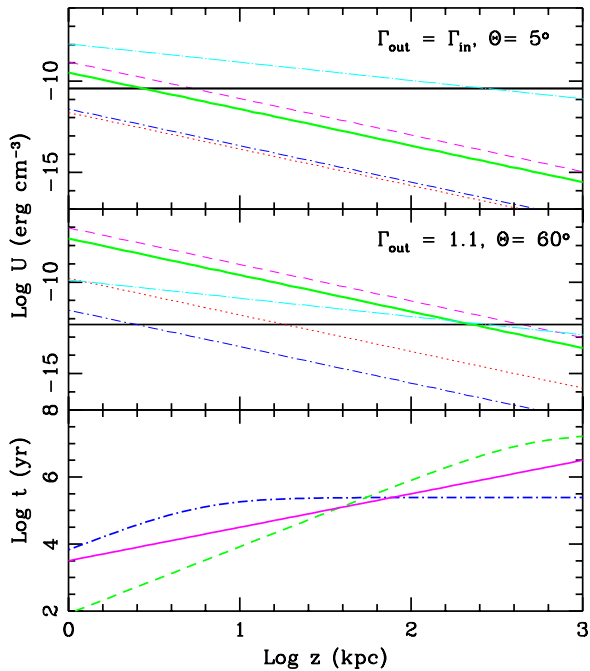,width=0.55\textwidth,angle=0}}
\caption
{Energy densities as a function of $z$ for a highly relativistic knot 
($\Gamma_{\rm out} = \Gamma_{\rm in} \sim 10$) ({\it upper panel}) and 
a mildly relativistic one ($\Gamma_{\rm out} \sim 1.1$) 
({\it middle panel}).
We consider these two components to dominate at small and large
observing angles, respectively.  
The various fields are represented as: 
$U^\prime_{\rm in}$ (solid, oblique line), 
$U^\prime_{\rm B}$ (dashed), 
$U^\prime_{\rm kpc}$ (dotted), 
$U^\prime_{\rm s,out}$ (dot--dashed), 
$U^\prime_{\rm CMB}$ (solid, horizontal). 
We have assumed intrinsic luminosities 
$L_{\rm in}= 10^{43}$, 
$L_{\rm s,out}= 10^{42}$,
$L_{\rm B}= 10^{45}$ and
$fL_{\rm disk}=10^{43}$ erg s$^{-1}$, $T_{\rm dust}=20$ K and reshift zero.
The particle distribution is assumed to be a power law of slope
$p$=2.5 between $\gamma_{\rm max} \sim 10^5$ and $\gamma_{\rm min} \sim 100$, 
with a corresponding energy density $U_{\rm e}$ (dot-long-dash line).  
The {\it bottom panel} shows the radiative cooling timescales 
(dash--dot line for the spine and dashed for the layer) 
compared with the timescale for adiabatic losses (solid),
calculated as $t_{\rm ad}=0.1 z/c$}
\end{figure}
Fig. 1 shows that along most of the jet 
the CMB is the dominant field for the spine: the energy densities of 
the magnetic field and of the nuclear blazar emission becomes
relatively important only in the inner parts ($<$ few kpc).
On the contrary for a slowly moving layer the beamed inner radiation 
can dominate the radiation energy density up to large scales. 
Reprocessed radiation from the NLR and dust
structures do not provide a relevant contribution at any distance.

For the spine emission, a systematic increase of the 
Compton dominance (ratio of the luminosity in the high energy and 
low energy spectral components) is expected with increasing distance. 
Clearly in terms of the observed luminosity the relative contributions 
of the spine and layer have to be weighted by the corresponding beaming
factors.

\subsection{Particle energy density}

Finally, let us consider the relative importance of the energy density
in relativistic electrons and fields (see Fig. 1). The electron
density can be estimated from the intrinsic synchrotron luminosity
$L^\prime_s=L_s/\delta^4_{\rm out}$ of the knot:
\begin{equation}
n^{\prime}_{\rm e} \, =\, 
{3 L^\prime_{\rm s, out} \over 4\pi R^3 c
\sigma_{\rm T} U'_{\rm B} \langle\gamma^2\rangle}.
\end{equation}
Then the electron energy density $U'_{\rm e} = n'_{\rm e}
\langle\gamma\rangle m_{\rm e} c^2$ is
\begin{equation}
U'_{\rm e} \, = \, 1.2\times 10^{-2} \, 
{U^\prime_{\rm out} \over U^\prime_{\rm B}} \,  {3-p \over p-2} \,
{( \gamma_{\rm max}/ \gamma_{\rm min})^{p-2} \over 
\psi_{-1} z_{\rm kpc}\gamma_{\rm max}} 
\quad {\rm erg \, \, cm^{-3} }
\end{equation}
where $\psi=10^{-1}\psi_{-1}$. 
We have assumed that the particle distribution is a
power law, $N(\gamma)\propto \gamma^{-p}$, between $\gamma_{\rm min}$ and
$\gamma_{\rm max}$, with $2<p<3$. 
For illustration, assume $p=2.5$, $\gamma_{\rm min}=100$, 
$\gamma_{\rm max}=10^5$, 
and equal synchrotron and magnetic energy densities.
In this case 
$U^\prime_{\rm e} \sim 4\times 10^{-6} \psi^{-1}_{-1} z^{-1}_{\rm kpc}$ 
erg cm$^{-3}$.
It is then clear that if $U^\prime_B\sim U^\prime_{\rm out}$ holds, 
then the energy density of relativistic electrons 
is bound to dominate by a large factor, especially at large distances,
as long as $\psi z$ increases with $z$ 
(since $U^\prime_{\rm e}$ scales as $z^{-1}\psi^{-1}$ 
while the other energy densities scale as $z^{-2}\psi^{-2}$).

The only way to decrease the particle dominance is to increase the relative
importance of $U^\prime_{\rm B}$.
But this immediately implies that the SSC component
of the spectrum becomes weaker than the synchrotron one.
If a large Compton X--ray flux is instead observed, then a more moderate
particle dominance can be achieved only if this emission is not due to SSC.
Equipartition between $U^\prime_{\rm B}$ and $U^\prime_{\rm e}$ is 
however possible if the external radiation fields are orders of magnitude 
greater that the local synchrotron energy density (if the X--ray flux is 
comparable to the low energy synchrotron one).
Such large ratios of external to local radiation fields are possible for 
the spine component (see Fig. 1) at large distances.
The spine component of jets can then have a Compton dominated spectrum
even in the case of equipartition between magnetic field and particles,
providing that the magnetic field in the emitting region is greater than 
the values shown in Fig. 1 (possibly due to shock compression 
and field amplification).

In summary, suppose to observe a significant inverse Compton 
X--ray flux in an extended jet structure.
If the source is a radio--galaxy, we have a strongly particle dominated 
emitting region, since we are likely observing the emission from the layer,
for which the external photon fields can only slightly reduce 
the particle to magnetic energy density ratio.
If the source is an extended jet of a blazar, we are observing 
the emission from the spine, with a strongly dominating CMB field.
In this case it is possible (even if not guaranteed), that
$U^\prime_{\rm B} \sim U^\prime_{\rm e}$.

\section{Multifrequency mapping}

In aligned sources (i.e. superluminal sources and blazars) the
more visible parts of the jet will be those still moving at
relativistic speeds, producing radiation highly beamed towards the
observer (i.e. the spine emission will dominate).  
On the contrary, in misaligned radio--galaxies the slower moving parts
(layer) dominate with quasi--isotropic radiation providing most of the
observed flux.

A further crucial point to be considered in such extended
structures is the relative role of radiative versus adiabatic cooling.
The bottom panel of Fig. 1 shows the radiative cooling timescales
of $\gamma=10^5$ electrons of both the spine and the layer.
Due to the different radiation fields dominating in the spine and in the 
layer, the spine is radiatively efficient in the outer 
parts and the layer in the inner parts.
Therefore at large distances the adiabatic losses are negligible for 
the spine, and are instead important for the layer.

In Fig. 2 we show examples of the SED for the spine and layer, 
where the corresponding jets are assumed to be
observed at $5^\circ$ and $60^\circ$ from the jet axis, respectively.  
The particle distribution is normalized to give a fixed
amount of {\it total} (synchrotron plus Compton) luminosity.
The addition of the external radiation field then
changes also the synchrotron flux level, and greatly changes
the Compton to synchrotron flux ratio.
This can be seen in the upper panel of Fig. 2, comparing the SED
corresponding to SSC only with the one including the contribution of
the external nuclear photons.
Note that the pure SSC case can be appropriate for a curved or bent jet, 
in which the beamed radiation from the center does not 
illuminate the large scale jet.

A further prediction of the model due to the energy dependent cooling
timescales is that the relative extension from the dissipation region
where particles are accelerated would be larger for slower cooling
particles, i.e. will depend on the mapping frequency, as $\propto
c t_{\rm cool} (1+z) \Gamma^{-1}$.

\begin{figure}
\vskip -0.5 true cm
\centerline{\psfig{figure=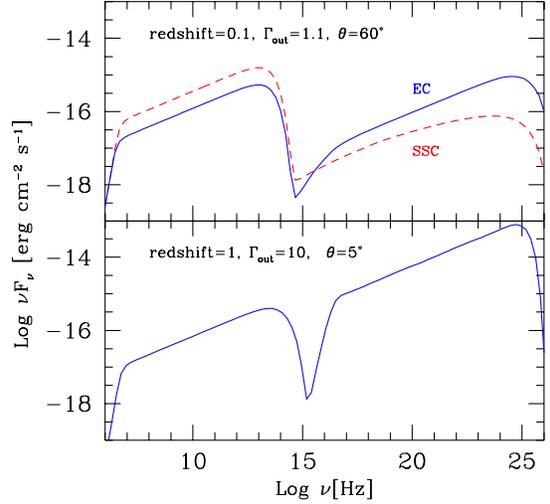,width=0.45\textwidth,angle=0}}
\vskip -0.5 true cm
\caption{
SED calculated assuming that at distances of 10 kpc from the center
a region of 1 kpc of size embedded in magnetic field
of $10^{-5}$ G radiates an intrinsic power of
$3\times 10^{41}$ erg s$^{-1}$.
The nuclear (blazar) component emits an intrinsic power of
$10^{43}$ erg s$^{-1}$. The power law electron distribution
has slope $p=2.5$ between $\gamma_{\rm min}=30$ and $\gamma_{\rm max}=10^6$.
The {\it upper panel} shows the case of the emission from a layer 
moving with $\Gamma_{\rm out}=1.1$ at a viewing angle $\theta=60^\circ$.
The dashed line corresponds to the SED assuming that electrons
emit SSC radiation only.
The solid line takes into account the radiation field, coming from the 
nucleus, illuminating the region.
The {\it bottom panel} shows the emission from a spine moving with
$\Gamma_{\rm out}=\Gamma_{\rm in}=10$
at a viewing angle $\theta=5^\circ$.
We have assumed a redshift equal to 0.1 for the layer
and equal to 1 for the spine, in order to have comparable
observed fluxes.
}
\end{figure}

\section{The jet of PKS 0637--752}

Let us consider the specific case of the jet
associated with PKS 0637--752, and in particular of its knot W7.8, for
which detailed information are reported (Chartas et al. 2000).

The spatially resolved X--ray images allowed to identify knots of
emission similar in structure and intensity to those seen in the radio
(Tingay et al. 1998) and optical (HST; Schwartz et al. 2000) bands.
According to the CXO and radio results, the source can be resolved at
5 GHz with a dimension of 0.3", at a distance from the central source
of about 7.8", yielding $\psi \sim 0.04$.
VSOP observations detected superluminal motion in the (pc
scale) jet (Lovell 2000), setting a lower limit on the (inner) bulk
Lorentz factor $\Gamma >17.5$ and an upper limit on the viewing angle
$\theta< 6.4$ degrees (assuming $H_0=50$ km $s^{-1}$ Mpc$^{-1}$ and
$q_0=0$).  The small scale jet appears to be well aligned with the kpc
scale one.

As discussed by Chartas et al. and Schwartz et al., the radio spectrum
can be explained as synchrotron emission, as also supported by the
detection of strong linear radio polarization, while the X--ray
emission largely ($\sim$ 2 orders of magnitude) exceeds the
extrapolation from the radio--optical spectrum.  
On the other hand, if the X--ray flux is due to  
SSC emission, large deviation from equipartition between magnetic field 
and particle energy density and/or strong inhomogeneities have to be 
invoked in the emitting region. 
Thermal emission by e.g. shocked plasma also appears
to be unlikely, as it would imply high particle density and thus
unobserved large rotation measures.
On the contrary in the proposed scenario the X--ray emission can be
satisfactorily interpreted as scattered CMB radiation.  

Fig. 3 shows the SED of both the
nuclear blazar component and of the knot W7.8.  
To fit the core emission, we have assumed an emitting region of size 
$R=9\times 10^{16}$ cm, magnetic field $B=4$ Gauss, bulk Lorentz factor
$\Gamma_{\rm in}=17$ and viewing angle $\theta=6^\circ$ 
(yielding $\delta_{\rm in}=8.2$).  
The emitted intrinsic luminosity is $L^\prime= 8\times 10^{43}$ erg s$^{-1}$.  
For this nuclear jet component we could estimate the kinetic power carried by 
the jet: in
the form of Poynting flux and emitting electrons we have respectively
$L_{\rm B}=1.4\times 10^{46}$ erg s$^{-1}$ $L_{\rm e}=1.4\times 10^{45}$.  
Assuming one (cold) proton for each electron, the estimated
kinetic power is of order $L_{\rm kin}=3\times 10^{47}$ erg s$^{-1}$.

\begin{figure}
\vskip -0.5 true cm
\centerline{\psfig{figure=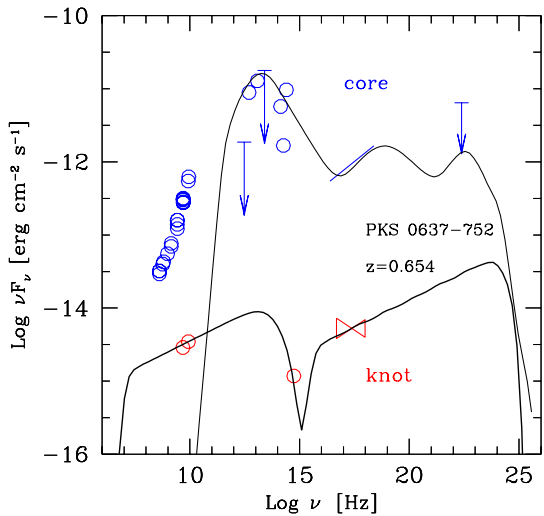,width=0.45\textwidth,angle=0}}
\vskip -0.5 true cm
\caption{The SED of the core and of the extended 7.8" knot
of the blazar PKS 0637--752.
Data sources: NED for the core;  Chartas et al. 2000 and
Schwartz et al. 2000 for the knot in the jet.
The solid lines are models calculated as explained in the text.
}
\end{figure}

For the knot, we have assumed that the emission region has a size
of $\sim$3 kpc, at a distance of 640 kpc from the center 
(corresponding to the deprojected observed distance, with a 
viewing angle of $6^\circ$).
The magnetic field is $B=6\times 10^{-5}$ G, the intrinsic 
luminosity is $L=10^{41}$ erg s$^{-1}$ and $\Gamma_{\rm out}=14$. 
The viewing angle is the same as the core.
With these parameters, the CMB contribution to the radiation fields 
as seen by the knot is largely dominating.
At the redshift of PKS 0637--752 this
corresponds to $U^\prime_{\rm CMB} \sim 6\times 10^{-10}
(\Gamma_{\rm out}/14)^2$ erg cm$^{-3}$, to be compared
with $U^\prime_{\rm B}\sim 1.4\times 10^{-10}$ and 
$U^\prime_{\rm out}\sim 2\times 10^{-15}$ erg cm$^{-3}$.
The jet bulk kinetic powers are 
$L_{\rm e}=5\times 10^{45}$ and
$L_{\rm B}=2.6\times 10^{47}$ erg s$^{-1}$. 
If there is one proton for emitting electron 
$L_{\rm kin}=4\times 10^{47}$ erg s$^{-1}$. 
These numbers, typical for powerful radio sources, 
are in rough agreement with the corresponding values for the core,
suggesting that the kinetic power is conserved along the jet
(see e.g. Celotti \& Fabian 1993).
Note that both the large bulk Lorentz factor and
the small size (i.e. a much smaller value of $\psi$ than
what adopted in the previous section) refer to the {\it spine} 
component, while the jet could appear wider and slower if the  
layer dominates (i.e. at large viewing angles).
However, so far direct evidence for layers exists only for 
$\sim$kpc scale jets (Swain, Bridle \& Baum 1998; Laing et al. 1999).

The optical emission can be self--consistently interpreted as
the higher energy part of the synchrotron component. This in turn
implies that the emission is due to high energy electrons, whose
cooling length corresponds to distances $\sim 0.5$ kpc. 
Radio and
X--ray emission are instead accounted for by particles of similar
(lower) energies and therefore images in these two bands are expected
to resemble each other (with cooling lengths $\sim 150$ kpc), as
indeed observed (Schwartz et al. 2000).

In our model, $\gamma_{\rm min}$ is required to be small (10--20) 
in order to fit the X--ray flux and spectrum.
The presence of such low energy electrons can cause some
depolarization of the radio emission of the knot (Wardle et al. 1998)
possibly indicating the contribution of electron--positron pairs
(which do not affect polarization).


\section{Discussion and conclusions}

We have discussed the role of different electromagnetic fields for the
radiative dissipation of relativistically accelerated electrons in
large scale jets. The dissipation regions (knots) act as scattering
mirrors for any externally produced photon field as well as the
nuclear beamed radiation. A jet structure comprising components with
different speeds can account for large scale emission in jets
associated with both blazar and radio--galaxies.

In the former case, we expect to detect more intense radiation from a
fast moving component (spine), spatially concentrated in the outer jet
regions and dominated by the CMB scattered field at distances greater than a 
few kpc. 
On the contrary, weaker and more isotropic emission would
dominate in radio--galaxies, produced by a mildly relativistic
portion of the jet, mainly Compton scattering the
beamed radiation coming from the nucleus.
Deviation from this behavior in radio galaxies is expected for jets
which are subject to strong bending: in this case the inner beamed
field would not reach any misdirected component.


While the above estimates imply values of the kinetic power associated
with the emitting knots typical of high power radio sources, a
quasi--stationary hydrodynamical confinement of such structures
appears to be a problem as extreme high gas densities for the external
gas would be required.

We expect that optical and X--ray emission to be common in
large scale jets of both blazar and radio--galaxies. The promising
results by $Chandra$ and HST will thus play a key role in the
understanding of the jet structure and dynamics as well as the
dissipation processes.

\vskip 0.25 true cm
\noindent
{\bf Acknowledgments}

\noindent
We thank the anonimous referee for useful suggestions.
This research has made use of the NASA/IPAC Extragalactic Database (NED)
which is operated by the Jet Propulsion Laboratory, Caltech, under
contract with the NASA.

\end{document}